# Identifying and understanding obstacles to heating sobriety and thermal comfort in collective housing: Insights from a survey in France

Enzo Cabezas-Rivière[1,2*], Maxime Robillart[1,2], Aline Barlet[3], Thomas Recht[1,2], Laurent Mora[1,2], Patrick Sebastian[1,2]

[1] Univ. Bordeaux, CNRS, Bordeaux INP, I2M, UMR 5295, F-33400, Talence, France
[2] Arts et Metiers Institute of Technology, CNRS, Bordeaux INP, I2M, UMR 5295, F-33400 Talence, France
[3] GRECCAU, ENSAP Bordeaux, 740 Cours de la Libération - CS70109, 33405, Talence, France

*Corresponding author* : **enzo.cabezas@u-bordeaux.fr**

*Permanent address* : I2M - UMR 5295, Université de Bordeaux - Bât A11, 351 Cours de la Libération, 33405 TALENCE CEDEX

**Abstract**:

In many countries, central heating systems are widely used in multifamily housing allowing maintenance and costs to be shared. However, these systems often limit residents' control over their own consumption, complicating efforts to reduce energy use and maintain comfort. Despite the growing importance of household energy savings in national and European climate targets, little research has examined user experiences with these systems. This study addresses that gap through an exploratory survey conducted in France.

Grounded in scientific literature, technical analyses, and current policy frameworks, the survey was distributed through various channels—including public institutions, online platforms, and field outreach—and collected 403 responses. It examined user difficulties in managing heating, including challenges with controls, bill comprehension, and communication with landlords. It also explored behaviours with high energy-savings impact: reducing heating at night or during absences, window opening, and shutter use.

Findings show that many residents face barriers to energy-efficient heating, often due to difficult-to-use controls, limited feedback on consumption, and poor support from landlords or property managers. Those who do not reduce heating at night or when away frequently report these challenges. Participants expressed strong interest in solutions such as personalized advice, real-time feedback, and connected radiator valves to improve comfort, control, and energy management.

This work highlights the need to better integrate user experience into multifamily housing energy policies. It provides actionable insights for policymakers and housing providers seeking to enhance comfort and energy savings while supporting France and Europe's broader energy transition goals.

# 1. Introduction

Worldwide, building energy consumption accounts for more than 30% of the total final global energy use and contributes to approximately 28% of global greenhouse gas emissions (1).This substantial impact presents significant challenges for both environmental sustainability and economic stability. Within the overall energy balance of buildings, approximately 50% is devoted to space and water heating, resulting in an estimated 2,400 Mt of direct and 1,700 Mt of indirect $CO_2$ emissions (2). A common strategy involves improving building energy efficiency, such as through renovations. However, studies show a persistent gap between projected and actual energy savings, even after such improvements (3–5). This gap is partly due to the fact that energy performance depends not only on the building envelope and systems but also on occupant behaviour. Behavioural changes alone can yield up to 30% in energy savings (6). A purely technical approach risks triggering a rebound effect, where expected savings are offset by stable or increased consumption (6–8). Yet behaviour change is complex, shaped by diverse and interacting factors (9–11) including:

- **Psychosocial factors**, including age (12–16), gender (16), education level (17) motivation to save energy (18–20), social norms (21–23) and psychological processes related to environmental awareness (9,13,14,23–25);
- **External or contextual factors**, which can facilitate or hinder the adoption of behaviours. In the case of heating in buildings it could be the building and/or the system characteristics that occupant use, such as the type of heating system (12,16,26), type of heating control (14,17,27,28);
- **Psychophysiological factors,** refer to the ways in which individuals seek to achieve and maintain comfort, including for example thermal comfort (11,29–31), acoustic comfort (30) (32) or visual comfort (30,32). Comfort itself is influenced by environmental factors such as temperature, light, noise levels and air quality.

These factors influence energy-saving behaviours through distinct but interacting mechanisms. Psychosocial factors, such as knowledge, attitudes, social norms, and motivation, shape how occupants perceive energy use and decide whether to adopt energy-saving practices (23,33). For instance, occupants who are more aware of environmental impacts or motivated to reduce costs may adjust their heating use more actively. External and contextual factors, including building and system characteristics, determine the scope and feasibility of behavioural adjustments. Constraints such as the type of heating system, control interfaces, or limitations on modifying equipment can restrict occupants' ability to act on their intentions, regardless of motivation or awareness (10,34). Psychophysiological factors act as a key driver of heating behaviour, as individuals tend, whenever possible, to adjust their use in order to achieve and maintain a desired level of comfort. Together, these mechanisms help explain why energy use often deviates from predicted levels and highlight the importance of considering behavioural, technical, and contextual factors when assessing heating performance.

To promote behavioural change, interventions are commonly employed. In the context of energy-saving behaviours in buildings, interventions include, for example, persuasive messages (35,36), feedback on energy consumption (37–39) and energy-saving goals (40,41). To improve the robustness and relevance of interventions, studies suggest that behavioural change factors should be considered when selecting interventions (42,43). In this context, recent studies have proposed methods for tailoring interventions to individuals' psychosocial profiles (44,45). While psychosocial and psychophysiological factors involved in behaviour changes are relatively well understood, this is not the case for external or contextual factors that may hinder the adoption and maintenance of energy-saving behaviours (43).

This study focuses specifically on these contextual factors related to collective heating systems in France, such as cost allocation methods, the collective nature of heating consumption, and where the regulations in place may be insufficient to fully leverage behavioural drivers for energy savings (46). The motivation behind this work is to identify the limiting factors in such systems and to propose tailored interventions that can induce and guide energy savings while maintaining occupant comfort.

Collective (or central collective) heating refers to a system in which heat is generated from a single source and distributed to multiple units or dwellings within a building or group of buildings, typically for space and/or water heating. In Europe, these systems account for 15–20% of heating installations (47). In buildings with collective heating, costs are typically distributed according to apartment size, meaning two similarly sized units pay the same amount, regardless of actual usage. This method disregards behavioural differences: residents who adopt energy-saving habits pay as much as those who do not. Consequently, individual efforts may seem useless, as reducing the heating bill requires collective participation (48,49). To address this issue, since 2012 the European Energy Efficiency Directive introduced the objective of individualised heating cost allocation in collective heating systems (50). By linking charges to actual consumption, this individualisation increases users' awareness of their heating use and could lead to energy savings of around 15% (51). This raises a broader question: Do occupants of residential buildings equipped with collective heating systems face difficulties or obstacles in saving energy and maintaining comfortable living space? In the literature, very few studies focus specifically on collective heating systems in residential buildings and the difficulties individuals may face in saving energy while maintaining a satisfactory level of comfort.

Yin et al. (52) found that in 12 residential buildings in Beijing, collective central heating often led to overheating in certain dwellings, resulting in thermal discomfort for occupants and that this system limits their ability to regulate indoor temperatures according to their personal preferences. Findings from a French study (53) show that measured indoor temperatures tend to be higher in dwellings with collective heating compared to those with individual heating. However, the reasons for this difference may be multiple, including overheating, lack of user control over heating settings, or low awareness of energy-saving practices. Discomfort situations are also reported in national studies from France (54–56), where residents of collective housing appear to experience more thermal discomfort than those living in other types of residential buildings, often due to technical issues with heating systems, humidity problems, or poor thermal insulation.

Through a survey conducted across several European countries with 10,000 respondents, Sovacool et al. (57) have shown that the information available about energy consumption to individuals in collective heating systems appears to hinder greater awareness of energy savings and, consequently, energy sobriety. However, the study was not specifically designed to investigate the issues related to this available information, leaving unclear exactly how it may be limiting. One possible hypothesis is that heating charges are often bundled with other costs—such as maintenance of common areas or elevator upkeep—with only the total amount in euros being communicated. This format prevents residents from identifying periods of high consumption or understanding how their behaviours influence the final bill (58,59).

Sovacool et al. (57), highlight also that users often struggle to operate their heating systems and express a need for more intuitive and user-friendly technologies. However, the specific aspects of the systems that are perceived as difficult to use are not clearly identified, leaving open questions about the exact nature of these usability challenges. One possible explanation in collective housing is the widespread use of water radiators equipped with manual or thermostatic valves (60). These devices often display setpoints on a numerical or symbolic scale (e.g., "+" to "−") rather than in degrees, which can make them difficult to interpret. Moreover, as they are neither remotely controllable nor programmable, users must adjust them manually each time they wake up or return home. This lack of usability may hinder the adoption and maintenance of energy-saving behaviours.

Those factors identified in the literature may hinder the adoption and persistence of energy-saving heating behaviours. Although comfort—particularly discomfort situations—has been relatively well studied in collective heating systems in France, further research remains necessary. While some studies have highlighted individual challenges faced by residents, to the best of our knowledge no study has comprehensively examined the full range of difficulties and their underlying causes experienced by occupants of buildings equipped with collective heating systems. Furthermore, no research has analysed how these obstacles affect occupant behaviours or evaluated tailored solutions.

In this context, this study first aims to determine whether residents in buildings with collective heating systems face barriers to energy saving, focusing particularly on challenges related to the availability and clarity of energy information as well as the usability of heating controls. Then, it seeks to assess how these obstacles influence energy-saving behaviours. Finally, the study explores targeted interventions designed to help overcome these difficulties. From these objectives, three research questions arise to guide the investigation:
- **RQ1:** Do residents with collective heating systems face obstacles in managing their heating consumption? If so, what specific difficulties do they encounter?
- **RQ2:** What is the relationship between the difficulties faced and the energy-saving behaviours adopted by residents?
- **RQ3:** Which solutions do residents consider relevant and effective for reducing energy consumption?

## 2. Materials and Methods

To explore these questions, a national questionnaire survey was conducted, using the French residential sector as a starting point, where collective heating systems serve around 18% of housing (61). This context provides a relevant setting to identify the main challenges residents face in managing heating consumption. Although focused on France, the methodology can be applied to other European countries with similar collective heating systems, such as Germany, Italy, and Sweden (47).

### 2.1. Research design and analytical protocol

The questionnaire was developed as part of a multi-sectoral approach, combining insights from academic, technical, and policy perspectives. First, a comprehensive review of the scientific literature informed the formulation of key hypotheses—such as the impact of billing format on user understanding, and the challenges posed by poorly designed or unintuitive heating systems. In parallel, a technical review of collective heating infrastructure in France (62) helped identify the most common system configurations, notably central hot-water systems using radiator-based distribution with either manual or thermostatic valves. Finally, a legal and regulatory review contextualized the proposed solutions within both current and forthcoming national frameworks—particularly those related to heating cost individualization and energy performance obligations. This triangulated approach ensured the questionnaire was well-grounded in academic theory, technical realities, and relevant policy developments.

An online, self-administered questionnaire was created using Sphinx IQ2© software. The survey was anonymous, voluntary, and no compensation was provided. It was structured into seven thematic sections, comprising a total of 67 questions, with an estimated completion time of approximately 11 minutes. To reduce respondent burden and ensure the relevance of questions, the survey employed conditional logic, meaning that not all participants answered every question. For example, those who reported no heating-related difficulties were not shown follow-up questions about potential causes, and respondents without numerical setpoint displays or external shutters were not asked about their usage habits. Collecting the structural characteristics of the home also allowed us to adapt questions to each respondent's situation and to explore potential relationships between housing features and energy-related behaviours or comfort. The structure of the questionnaire was as follows:
- **Section One** identified the structural characteristics of the respondent's home and their occupancy status (e.g., owners, tenants, housemates, or residents with free accommodation).
- **Section Two** focused on understanding the obstacles to better controlling heating consumption and comfort, obstacles to understanding information related to heating consumption, as well as social interactions between respondents and their /property manager (RQ1).
- **Section Three** focused on individual habits with the greatest impact on heating consumption—heating use, window opening, and external shutter operation—in order to examine their relationship with the difficulties identified in Section Two (RQ2).
- **Section Four** aimed to identify and better understand sources of discomfort within the home.
- **Section Five** assessed individuals' attitudes toward energy savings.
- **Section Six** evaluated the relevance of proposed solutions that could help individuals manage their energy consumption and comfort more effectively (RQ3).
- **Section Seven** gathered socio-demographic information from respondents and assessed their awareness of their heating bills.

The responses were collected between December 2022 and May 2023. 411 responses were obtained, but 8 were excluded following a quality check. This check involved identifying "flat-liners" (respondents who gave the same answers across multiple questions), "rushers" (individuals who provided incomplete, contradictory, or unrealistic responses, such as claiming to live in a 40-person household in a 10-square-meter home), and "speeders" (respondents who completed the survey in under 5 minutes). After applying these quality controls, the final sample consisted of 403 respondents. The databases, data dictionary, and questionnaire form are available for further information (63).

## 2.2. Sample selection

Precise socio-demographic information about individuals living in collective heating systems in France is scarce. Existing national housing surveys generally address the broader population and do not provide detailed data specific to the target population, who may differ significantly in terms of socio-economic characteristics—for example, social housing tenants often have lower incomes compared to private homeowners or tenants (64).

Given this lack of detailed information on the target population, this study is exploratory in nature and relies on a convenience sampling approach. Among the limited data available on this population, one rare source provides information on occupancy statuses indicating roughly 72% tenants, 24% owner-occupants, and 4% others (60). The aim was to gather as many responses as possible while maintaining an order of magnitude consistent with the distribution of occupancy statuses in collective housing.

To maximize reach, the survey was distributed via multiple channels, including mailings through public institutions such as universities, town halls, and social housing providers. Additionally, in-person visits were conducted in residential areas of Bordeaux—where the authors live and work—to engage directly with residents. Finally, the questionnaire was published on the online platform SurveyCircle (65), which facilitates cooperation among researchers seeking survey participants. Consequently, participation was limited to individuals with internet access via a computer, tablet, or smartphone who were exposed to our recruitment materials and chose to take part in the survey. This likely introduced a degree of selection bias, which should be considered when interpreting the findings.

## 2.3. Data analysis techniques

Given the exploratory nature of this study, descriptive methods were primarily employed to address the three research questions. Frequency analyses and cross-tabulations were used to summarize and visualize respondents' experiences and perceptions, notably including the characteristics of the study population, the difficulties encountered, the reported energy-saving behaviours, and the interventions considered relevant by participants to help reduce energy consumption without compromising comfort.

The analysis aimed to investigate whether the difficulties encountered by respondents were related to each other, and in particular, whether they were associated with reported energy-saving behaviours. To achieve this objective and identify links that could guide the prioritisation of improvements and the design of tailored interventions, statistical tests were used. Since the questionnaire generated predominantly categorical data, two statistical tests commonly used for analysing associations between categorical variables were employed: Chi-square ($\chi^2$) and Fisher's exact test (66,67). The choice between the two depended on the distribution of observations within contingency tables. Chi-square tests were used when all expected cell counts were greater than 5, ensuring the validity of the $\chi^2$ approximation, whereas Fisher's exact test was applied whenever at least one expected cell count fell below this threshold. Using Fisher's test in such cases helps avoid the overestimation of statistical significance that can occur with sparse data. Because Fisher's test does not produce a conventional test statistic, results are reported using the estimated odds ratio and its 95% confidence interval. Odds ratios greater than 1 indicate a positive association, whereas odds ratios below 1 indicate a negative association; confidence intervals that include 1 suggest that the association is not statistically significant. For $\chi^2$ tests, test statistics and corresponding p-values are reported. Lower p-values indicate stronger evidence against the null hypothesis: values below 0.05 are considered statistically significant, values below 0.01 indicate a stronger level of significance, and values below 0.001 reflect very strong evidence of an association. All analyses were conducted using the R statistical software (v. 2022.12.0+353).

## 3. Results

### 3.1. Sample Characteristics

Table 1 presents the socio-demographic characteristics of our study sample compared to the representative 2022 French population (68). Given the limited data available on the target population, particularly those living in collective housing, the only specific national breakdown concerns occupant status (tenant vs. owner). Therefore, we consider it relevant to compare our sample with national distributions for contextual purposes. This comparison is not intended to assess representativeness but to situate our findings within a broader framework.

Regarding occupant status, 78% of respondents are tenants, 13% are homeowners, and the remaining 9% live rent-free. While these figures differ from national averages (32% tenants and 64.7% homeowners), they are closer to the limited data available for residents in collective buildings. For instance, national statistics indicate that 73% of occupants in collective housing are tenants or live rent-free—comparable to the 87% observed in our sample—while 20.6% are homeowners, versus 13 in our data (64). The high proportion of tenants is also explained by the large presence of students, only 5% of whom own their homes.

Additionally, the survey sample includes a higher proportion of respondents from certain regions, particularly Nouvelle-Aquitaine and neighbouring Occitanie. The sample also contains a notable share of students, exceeding the 9.2% reported in the national Employment Survey (68). These features mainly reflect the survey's distribution method and the absence of targeted sampling by socio-professional category.

*Table 1: Socio-demographic characteristics of the study population, compared with the French population*

| Characteristics | Sample (n=403) | | French population from the national employment survey (68) (n=348 964) |
|---|---|---|---|
| | Number | Frequency (%) | Frequency (%) |
| **Gender** | | (n=403) | (n=348 964) |
| *Woman* | 225 | 56% | 53.1% |
| *Men* | 170 | 42% | 46.9% |
| *Does not wish to comment and Non-binary* | 8 | 2% | - |
| **Main situation** | | (n=403) | (n=343 269) |
| *Employed* | 142 | 35% | 45.6% |
| *Unemployed* | 9 | 2% | 6.3% |
| *Retired or pre-retired* | 5 | 1% | 30.7% |
| *Studying* | 239 | 60% | 9.2% |
| *Others & no answer* | 8 | 2% | 8.2% |
| **Socio-Professional Categories** (among "Employed" and "Unemployed") | | (n=151) | (n= 177 255 ) |
| *Managers and higher intellectual professions* | 76 | 50% | 20.6% |
| *Employees* | 43 | 29% | 27.0% |
| *Craftsmen, traders, business owners, manual workers and agricultural operators, and intermediate professions* | 23 | 15% | 52.4% |
| *Unemployed[1]* | 9 | 6% | - |
| **Housing occupancy status** | | (n=403) | (n=56 396) |
| *Tenants* | 314 | 78% | 32.0% |
| *Owners* | 54 | 13% | 64.7% |
| *Hosted free of charge* | 35 | 9% | 2.3% |
| *Usufructuary* | - | - | 1.0% |
| **Region of residence** | | (n=403) | (n=429 027) |
| *Nouvelle Aquitaine* | 104 | 26% | 8.4% |
| *Ile-de-France* | 103 | 26% | 16.0% |
| *Occitanie* | 52 | 13% | 6.6% |
| *Auvergne Rhône-Alpes* | 39 | 10% | 10.6% |
| *Grand Est* | 31 | 8% | 9.0% |
| *Hauts-de-France* | 18 | 4% | 8.1% |
| *Provence-Alpes Côte d'Azur* | 15 | 4% | 6.3% |
| *Pays de la Loire* | 14 | 3% | 5.0% |
| *Normandie* | 8 | 2% | 4.9% |
| *Bourgogne Franche-Comté* | 7 | 2% | 4.6% |
| *Centre-Val de Loire* | 6 | 1% | 3.4% |
| *Bretagne* | 6 | 1% | 4.8% |
| *Overseas departments and regions / Corsica* | - | - | 12.3% |

[1] Socio-professional category was not collected for respondents who were unemployed, whereas it is available for the French population in the National Employment Survey.

## 3.2. RQ1: Obstacles encountered by people with a collective heating system

Various propositions were presented to respondents to identify the difficulties they encountered. Figure 1 illustrates the difficulties reported by the participants. The primary difficulties are mainly related to heating bills, corresponding to the combined proportion of responses in the categories "strongly disagree", "disagree", and "somewhat disagree"):
- Understanding the distribution of the building's heating costs (54%).
- Estimating the impact of heating use on the bill (54%).
- Anticipating heating bills (50%).
- Understanding heating bills (44%).
- Accessing heating bills (41%).

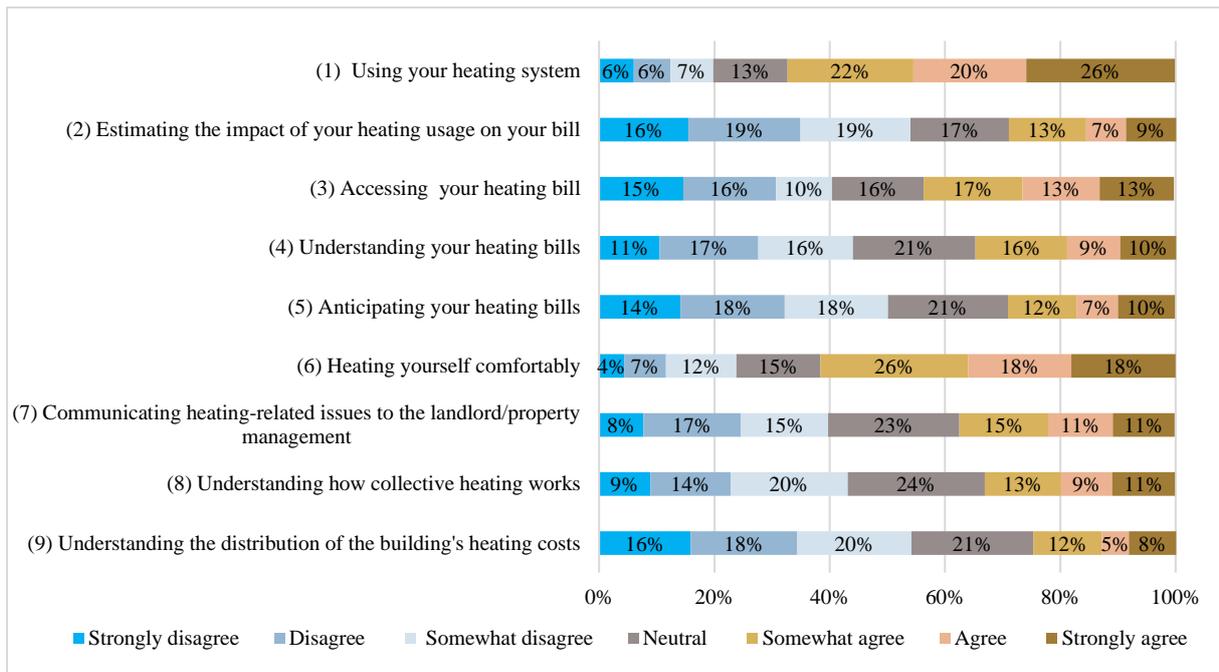

*Figure 1: Difficulties encountered by respondents. Answers to the question: "In your household, do you think it is easy to..." (n=403, scale question)*

The question "What difficulties do you encounter in adjusting your heating for comfortable use?" (n=200, multiple-choice) aimed to identify the causes of the difficulties faced in heating comfortably and/or using the heating system (seen in Figure 1). Results show that these difficulties are mainly caused by non-programmable radiators (60%), non-remotely controllable systems (39%), and excessively high bills (33%).

In response to the question "Do you know the annual amount of your heating bill?" (n=403, Yes/No question), only 12% of respondents answered yes—87% of non-owners and 83% of owners did not. Among those unaware of the amount, 32% cited limited access to bills, while 42% attributed this to heating costs being bundled with other charges.

Table 2 presents statistical associations between the difficulties highlighted in Figure 1. Column variables are labelled with numbers corresponding to the rows (e.g., (1) refers to difficulty in using the heating system). To facilitate statistical analyses, variables related to difficulties were recoded into three categories:
- No difficulty, grouping responses of "strongly agree," "agree," and "somewhat agree";
- Neutral, corresponding to the "neutral" response;
- Difficulty encountered, including "somewhat disagree," "disagree," and "strongly disagree."

*Table 2: Statistical cross-tabulations between the difficulties encountered by individuals (n=403).*
*<u>Legend</u>: The values shown correspond to the chi² (χ²) test statistics.*
*\*\*\* p < 0.001, \*\* p < 0.01, \* p < 0.05, and X indicates a statistically non-significant test.*

| Difficulties encountered (n=403) | (1) | (2) | (3) | (4) | (5) | (6) | (7) | (8) | (9) |
|---|---|---|---|---|---|---|---|---|---|
| (1) using their heating | - | 22.43 *** | 22.02 *** | X | 18.56 *** | 32.27 *** | 32.27 *** | 20.96 *** | X |
| (2) estimating the impact of usage on the bill | 22.43 *** | - | 118.26 *** | 132.78 *** | 114.91 *** | 13.41 ** | 11.50* | 33.13 *** | 75.99 *** |
| (3) accessing heating bills | 22.02 *** | 118.26 *** | - | 117.82 *** | 63.28 *** | X | X | 35.01 *** | 76.01 *** |
| (4) understanding heating bills | X | 132.78 *** | 117.82 *** | - | 119.48 *** | 11.69* | 11.35* | 83.19 *** | 111.87 *** |
| (5) anticipating heating bills | 18.56 *** | 114.91 *** | 63.28 *** | 119.48 *** | - | 23.17 *** | 33.75 *** | 31.69 *** | 97.82 *** |
| (6) heating comfortably | 75.17 *** | 13.41 ** | X | 11.69* | 23.17 *** | - | 47.13 *** | 18.07 ** | X |
| (7) communicating with their landlord/property manager about heating-related issues | 32.27 *** | 11.50* | X | 11.35* | 33.75 *** | 47.13 *** | - | 14.77 ** | 22.42 *** |
| (8) understanding how collective heating works | 20.96 *** | 33.13 *** | 35.01 *** | 83.19 *** | 31.69 *** | 18.07 ** | 14.77 ** | - | 103.63 *** |
| (9) understanding the distribution of heating costs | X | 75.99 *** | 76.01 *** | 111.87 *** | 97.82 *** | X | 22.42 *** | 103.63 *** | - |

☐ *** p < 0.001, ☐ ** p < 0.01, ☐ * p < 0.05, ☐ X (non-significant)

A significant proportion of the statistical tests indicate that individuals reporting one difficulty are likely to experience several others. The results show significant associations in most cases, except for:
- The difficulty in accessing heating bills (3) is not significantly correlated with difficulties in heating comfortably (6) or communicating heating issues to owners/property managers (7).

The difficulty in heating comfortably (6) is not significantly correlated with understanding the allocation of heating costs (9).

### 3.3. RQ2: Impact of obstacles encountered on declared energy-saving behaviour

Table 3 presents the heating usage habits reported by respondents. Firstly, the findings indicate that 39% of respondents lack the means to adjust their heating, while 4% are unsure if they have the means to do so. Among the 57% of respondents who have the capability to regulate their heating, 35% do not lower their heating when they are absent for a few hours.

*Table 3: Heating equipment and habits*

| Variables | n | Frequency (%) | | |
|---|---|---|---|---|
| | | Yes | No | Others |
| Having radiators with adjustable valves | 403 | **229** (57%) | 159 (39%) | 15 (4%) |
| Reducing heating during a few hours of absence | **229** | 139 (61%) | 81 (35%) | 9 (4%) |
| Adjusting radiators differently depending on the rooms in the home | 214 | 132 (62%) | 52 (24%) | 30 (14%) |
| Reducing heating at night in unoccupied rooms | | 104 (49%) | 83 (39%) | 27 (13%) |

Focusing on individuals with the ability to adjust their heating in both day-use areas (e.g., living rooms, dining rooms) and night-use areas (e.g., bedrooms, bathrooms), 24% report not adjusting radiator settings between different spaces, and 39% do not implement night-time setbacks. It is important to note that, as these responses are self-reported, they may be subject to response bias, including the potential overestimation of energy-saving behaviours. (9,69).

To investigate potential associations between reported difficulties and energy-saving behaviours, statistical analyses were conducted prior to the detailed presentation of the variables. The results indicate that respondents who do not reduce their heating during absences or at night are significantly more likely to report obstacles in managing their heating. Specifically, 49% of those who do not lower the temperature in either context report difficulties accessing their heating bills. Additionally, among individuals who do not reduce the temperature during absences, 52% report communication difficulties with their landlord property manager regarding heating. These associations were statistically significant. Not lowering the temperature during absences was associated with reduced odds of effectively accessing heating bills (Odd Ratio=0.35, Confidence Interval=[0.18 – 0.68], $p < 0.001$), and of communicating effectively with housing management (Odd Ratio=0.41, Confidence Interval=[0.20 – 0.81], $p < 0.01$). Similarly, not lowering the temperature at night was significantly associated with difficulties accessing heating bill information (Odd Ratio=0.39, Confidence Interval=[0.20 – 0.78], $p < 0.01$). Findings on differentiated heating between day and night-use areas align with those. It is important to note that for these statistical cross-tabulations, the "Neutral" category resulting from the recoding was excluded in order to retain only respondents who explicitly indicated whether or not they encountered a difficulty.

### 3.4. RQ3: Solutions to help occupants save energy
#### 3.4.1. Occupants' ideas and information judged essential for further energy savings

For residents who reported having control over their heating system (n = 157), an open-ended question was posed: *"In your opinion, what could help you save even more on your heating bills?"* Content analysis of the responses revealed that the most frequently mentioned terms were *bill* (14 occurrences), *radiator* (13), *double glazing* (9), and *thermostat* (8).

Detailed analysis reveals that respondents seek more accessible and understandable heating bills. Regarding "radiator", many expressed a desire to replace, upgrade, or gain remote or programmable control. For "thermostat", respondents emphasized the importance of remote control and programming capabilities. Additionally, respondents referred to measures related to energy efficiency, most notably a preference for double glazing as well as the replacement or improvement of collective boiler heating regulation systems. Fewer than 5% of responses suggested behavioural changes or increased energy-saving practices.

When asked, "What information would be essential for you to adjust your heating usage?" (n = 403, multiple-choice), 62% of respondents indicated a preference for access to data on energy savings. Additionally, 45% expressed interest in receiving personalised advice on how to reduce their heating consumption, 40% wanted the ability to view temperature settings in degrees, and 34% wished to know the $CO_2$ emissions generated by their heating usage. The desire to view temperature in degrees appeared largely independent of respondents' ability to adjust their heating, as it was equally common among those with and without adjustable radiator valves. Notably, 73% of respondents selected no more than two of the proposed options, suggesting a targeted or selective approach to the type of information they value. Conversely, 7% of respondents expressed no interest in the four proposed types of information, while 4% requested other types not included in the list. Of the respondents who specified, three wanted to manage their heating directly, and the others asked for clearer bills and/or real-time consumption data.

### 3.4.2. Assessing the relevance of proposals to help individuals better control their heating consumption

A set of proposals aimed at helping households save energy was presented to respondents (see Figure 2). On average, all proposals were deemed relevant, with none significantly standing out. While these results align with earlier findings (Figure 1), they may be influenced by social desirability bias, likely stronger for this question. Whether or not respondents reported specific difficulties did not appear to influence how relevant they considered the proposed measures. For instance, 85% of those who struggled to understand their heating bills and 86% of those who did not nevertheless found improvements to bill clarity relevant. Similar patterns were observed for difficulties related to bill consultation.

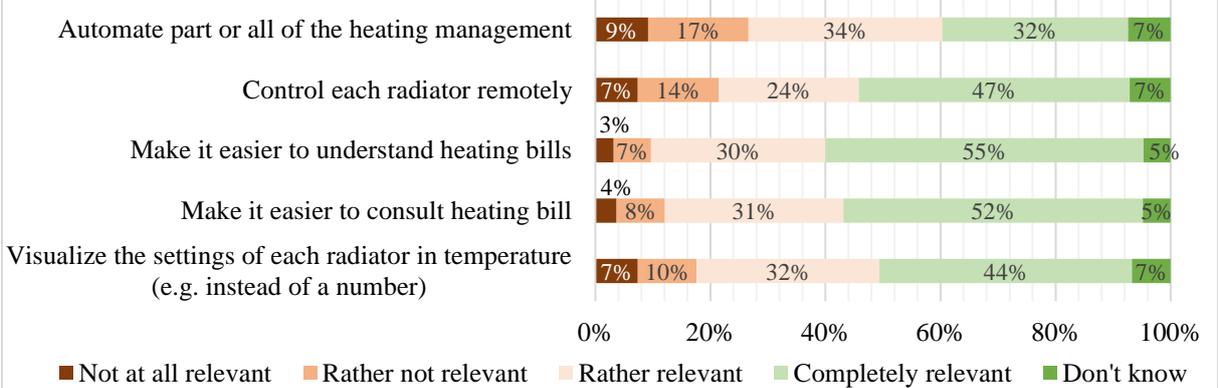

*Figure 2 : Evaluation of proposals aimed at helping individuals better adjust their heating usage. Answers to the question: "Which of the following suggestions would be most relevant for helping you further adjust your heating usage?" (n=403, scale question)*

## 4. Discussion

### 4.1. RQ1: Obstacles faced by households with collective heating systems

The findings clearly show that households in buildings with collective heating systems face multiple, interconnected difficulties in managing their heating use. These obstacles can be categorized as follows:

- **Low awareness of actual consumption and costs:** Many residents either do not have direct access to their heating bills, find them difficult to understand, or face bundled heating costs with other charges, which further distances them from their consumption patterns and their financial implications. This prevents them from effectively monitoring or adjusting their energy use.
- **Limited user control over heating systems**: Technical constraints, such as non-programmable or non-remotely controllable radiators, reduce residents' capacity to adapt heating to their needs.

- **Interdependence of difficulties**: Respondents who report one type of difficulty often experience several others. For example, lack of access to bills often coincides with difficulty interpreting costs or anticipating future charges, creating a reinforcing cycle of disengagement.

The systemic nature of these obstacles helps explain the limited prevalence of behavioural solutions reported by participants. Even motivated households may struggle to translate intentions into action when structural and informational barriers are present simultaneously.

Addressing these challenges requires a holistic approach that combines technical improvements, such as installing programmable or remotely controllable heating devices, with clearer, user-centred communication and transparent billing practices, thereby increasing the likelihood that interventions will support energy-saving behaviours while preserving occupant comfort.

## 4.2. RQ2: Link between difficulties and energy-saving behaviours

The data suggest a clear association between the difficulties reported by residents and their heating behaviours, indicating that structural and informational constraints play a central role in limiting action. Limited access to comprehensible billing information and insufficient communication with landlords or property managers appear to reduce the likelihood of engaging in basic energy-saving practices. While the direction and causality of these relationships remain uncertain, these patterns highlight that residents' inaction often reflects the systemic conditions in which they operate rather than a lack of interest or motivation.

No significant statistical links were found between reported difficulties and other potentially energy-relevant behaviours, such as window opening or shutter use. This suggests that the influence of structural or informational barriers is more pronounced in domains where control is either constrained or perceived as complex, such as managing indoor temperature through radiators.

Overall, these findings highlight that behavioural inaction is not merely a result of disinterest or disengagement, but is strongly influenced by the contextual conditions in which residents operate (70,71). When users lack access to reliable information or face uncertainty about the effects of their actions, they are less likely to feel empowered to change their habits. Consequently, efforts to promote energy efficiency must go beyond informational campaigns and consider the systemic and institutional obstacles that residents face in collective housing contexts.

## 4.3. RQ3: Solutions for reducing energy consumption

### 4.3.1. Individuals' perceptions of relevant solutions

Respondents consistently prioritized technical and informational improvements as key solutions to reduce energy consumption, emphasizing both enhanced control over heating systems—through programmable or remotely controllable radiators and thermostats—and broader energy efficiency measures, such as double glazing or improved boiler regulation. Interestingly, behavioural changes were rarely suggested in open-ended responses, which may reflect residents' difficulty in estimating the impact of their own actions on heating bills and the perceived complexity of influencing consumption within collective systems.

Respondents expressed a strong interest in energy savings data and personalized advice, regardless of whether they currently have the ability to adjust their heating. This points to a widespread demand for actionable, easy-to-understand information to support energy management.

While all proposed solutions were generally viewed as relevant, this was true even among respondents who do not experience the specific difficulties these solutions target. This collective endorsement highlights broad openness to interventions but may also reflect social desirability effects.

4.3.2. Our recommendations for overcoming identified difficulties

Based on the obstacles identified in collective heating systems, several targeted recommendations could be proposed to improve energy management and user engagement.

*4.3.2.1. Simplifying the use of heating*

With the growing accessibility of smart devices, installing connected radiator valves represents an affordable and effective solution. These systems provide remote control and scheduling, display setpoint temperatures in degrees, and offer real-time thermal comfort data (e.g., temperature and humidity), helping users better understand and regulate their heating use. However, in most cases in collectively heated buildings, occupants cannot install or modify heating control equipment themselves—such upgrades must be initiated by landlords or property managers. Installing connected valves in place of traditional or manual taps would not only align with regulatory requirements on the individualisation of heating costs (72), which mandate user control, but could also promote sustained energy-saving behaviour by making control easier and more intuitive.

However, it is relevant to question the environmental impact of such installations. A recent study assessed the footprint of connected devices in existing buildings using a life cycle analysis (73). The results show that the environmental impacts associated with deploying sensors for monitoring and control—as well as the energy needed to transmit and store data— are negligible compared to those from the building's use phase, particularly heating consumption. While further research is warranted, these initial findings underscore the value of such instrumentation in reducing both energy use and environmental impact.

*4.3.2.2. Enhancing the energy information available*

Enhancing the energy information available to occupants is essential to encourage energy-saving behaviours and enable residents to manage their heating knowledgeably. This involves providing clear, relevant, and actionable information that helps residents understand the impact of their actions on consumption and bills.

In this context, the obligation introduced in France on January 1, 2022, (74) , which requires monthly transmission of heating consumption data in buildings equipped with remotely readable meters, represents a timely opportunity. Beyond supplying residents with simple monthly consumption figures, this regulatory framework could be leveraged to provide more detailed, comprehensible, and actionable information, thereby enhancing occupants' ability to manage their heating systems effectively and adopt energy-saving behaviours. Improving energy information may include:

- Providing personalized advice based on thermal comfort data and user-set temperatures.
- Highlighting the impact of individual behaviours on energy use, with a focus on both actual and potential savings.
- Simplifying billing and cost allocation to reduce confusion and improve engagement.

For these improvements to be effective, the information provided must be tailored to occupants' actual practices rather than based on generic assumptions. Advice tailored to occupants' actual behaviours tends to be more engaging and perceived as relevant (75,76), which in turn increases its potential to influence both beliefs and actions (77,78). In the absence of individualised data, only general estimates can be offered, which may be perceived as inaccurate or irrelevant, potentially undermining both credibility and effectiveness (75,76). Moreover, self-reported data are often burdensome and susceptible to biases, including social desirability, highlighting the importance of objective behavioural measurements (79,80).

In this regard, data collected from connected radiator valves, as discussed in the previous section, offer a practical solution. Such data can support personalised, room-level recommendations, while also enabling the development of tools to estimate both potential energy savings and the actual reductions achieved following behavioural changes.

Indeed, accurately estimating the impact of user behaviour on heating consumption remains scientifically challenging, as heating demand results from a complex interplay between environmental conditions, indoor activities, and the building's physical characteristics. Addressing these complexities requires the development and calibration of dynamic thermal simulation models at the dwelling level (81). Such models can estimate heating consumption under varying usage scenarios and, importantly, make it possible to isolate the impact of changes in behaviours on consumption. This enables the quantification of both the financial savings and the reduction in environmental impact resulting from the energy-saving behaviours adopted. Despite advances in this area, accurately replicating real building behaviour remains a challenging scientific problem (82). These simulations could typically be conducted by specialised energy service providers, who might also manage heating cost allocation and could therefore already have dedicated models in place, as is the case for companies currently operating in the sector. At the scale of a single building or a few residential complexes, such providers could potentially leverage connected device data to generate personalised, room-level energy-saving recommendations for occupants, making the approach operationally feasible and potentially directly beneficial.

At a larger scale, it could be envisaged to conduct preliminary simulations across the main typologies of collective residential buildings in France, taking into account different building types, occupant behaviours, and even the position of dwellings within the building. Such an approach could allow the creation of a representative database, providing estimates of potential savings that more closely reflect real-world conditions. By referencing this database, it might then be possible to assess ranges of energy savings associated with behavioural changes rather than relying on single-point values, thereby offering more robust and generalisable insights into the likely effectiveness of energy-saving strategies.

4.4. Limits and perspectives of the study

This study has several limitations that should be considered when interpreting the results. Its exploratory nature required the use of convenience sampling, which may introduce selection bias and limit the generalizability of findings to the broader population of residents with collective heating systems. The reliance on self-reported data also raises concerns about response biases, such as social desirability or recall inaccuracies. Additionally, because the study focuses on the French context, the findings may not be directly applicable to other countries with different heating infrastructures or cultural practices related to energy use.

With a larger and more homogeneous sample—meaning a sample balanced in terms of socio-professional categories, residential locations, and heating system types—logistic regression analyses would be valuable to explore causal relationships between the variables identified as correlated in this study. They would also enable the identification of respondent profiles based on the difficulties encountered and the relevant interventions to support energy-saving behaviours. This would provide an evidence to inform and enhance policies aimed at encouraging behaviour changes in collective heating settings in France.

Building on this exploratory work, the next step is to conduct a more focused explanatory study with a larger and more homogeneous panel. This would enable a deeper understanding of how difficulties relate to socio-demographic characteristics, types of heating systems, and user habits. It would also support the identification of user profiles and the evaluation of targeted interventions best suited to help residents save energy effectively within collective heating contexts.

# 5. Conclusion

This study addressed a gap in the literature by comprehensively exploring the obstacles faced by residents interacting with collective heating systems, their impact on energy-saving behaviours, and residents' views on relevant solutions.

Our findings highlight major barriers to efficient heating, including difficult-to-use controls, unclear or inaccessible energy consumption information, and poor communication with landlords or property managers. These challenges make it harder for residents to understand bills, anticipate costs, and link their behaviour to energy use, reducing engagement in energy-saving practices. Respondents showed strong interest in personalized advice and real-time feedback, emphasizing the need for user-friendly controls and actionable energy information.

Two key recommendations emerge. First, installing connected radiator valves can simplify heating management by providing remote access, intuitive settings, and real-time feedback. Since residents often cannot install these themselves, landlords or property managers should lead the upgrades. These devices also support individualized heating cost regulations and deliver energy savings with minimal environmental impact.

Second, improving communication between residents and landlords or managers is essential. Clear, responsive channels can resolve issues, build trust, and provide personalized guidance based on real comfort data. Communication should clarify how behaviours affect energy use and simplify billing. Combining connected devices with dynamic thermal models can quantify behavioural impacts and support sustained energy-efficient behaviours. The recent French mandate for monthly heating data transmission offers an opportunity to enhance feedback with detailed, actionable insights.

Our contribution lies in bridging the existing knowledge gap by systematically identifying and analysing these interconnected difficulties within collective heating contexts. This work provides empirical evidence that effective energy efficiency strategies must address not only infrastructure but also the everyday realities and informational needs of residents. In doing so, the study contributes not only to academic understanding but also provides a practical foundation for informing public policy. By highlighting the importance of user-centered design, accessible information, and behavioural support, it offers valuable guidance for refining national and European policies aimed at decarbonizing residential heating. These findings support the development of more inclusive and effective policy instruments—particularly in the context of France's building renovation strategy and the European Green Deal—by aligning technological solutions with the real needs and constraints of end users.

Looking ahead, future research should build on these insights by incorporating richer demographic data and exploring social dynamics—such as -tenant relationships and neighbourhood conflicts—that may hinder energy-saving behaviours. It will also be important to link difficulty profiles to specific heating systems and, with larger and more homogeneous samples, identify resident profiles based on the challenges they face and the interventions they find most relevant. These analyses should be revisited in future studies to better establish causal relationships and clarify which specific barriers are most strongly associated with lower adoption of energy-efficient behaviours. Testing the practical implementation of tailored solutions will be essential to developing inclusive, effective strategies that empower residents, enhance comfort, and support a sustainable energy transition in collective housing.

# Declaration of generative AI and AI-assisted technologies in the writing process.

During the preparation of this work the authors used GPT-4 in order to improve the readability and language of the manuscript. After using this tool, the authors reviewed and edited the content as needed and take full responsibility for the content of the published article.

## CRediT authorship contribution statement

**Enzo Cabezas-Rivière**: Writing – review & editing, Writing – original draft, Methodology, Investigation, Formal analysis, Data curation, Conceptualization, Visualization. **Maxime Robillart**: Writing – review & editing, Validation, Supervision, Project administration, Methodology, Conceptualization, Funding acquisition. **Aline Barlet**: Writing – review & editing, Validation, Supervision, Project administration, Methodology, Conceptualization, Funding acquisition. **Thomas Recht**: Writing – review & editing, Validation, Supervision, Project administration, Methodology, Conceptualization, Funding acquisition. **Laurent Mora**: Writing – review & editing, Validation, Supervision, Methodology, Conceptualization. **Patrick Sebastian**: Writing – review & editing, Validation, Supervision, Project administration, Methodology, Conceptualization, Funding acquisition.

## Declaration of competing interest

The authors declare that they have no known competing financial interests or personal relationships that could have appeared to influence the work reported in this paper.

## Acknowledgments

This work was supported by a doctoral contract funded by the French Ministry of Higher Education and Research, awarded through the Doctoral School of Physics and Engineering of the University of Bordeaux.

## Data availability

The link to the data (including the supplementary material and citation on Mendeley Data) has been shared.